\documentclass[aps,prl,twocolumn,superscriptaddress,hidelinks]{revtex4-2}

\usepackage{color}
\usepackage{graphicx}
\usepackage{siunitx}
\usepackage{amsmath}
\usepackage{hyperref}

\begin{document}

\title{Vortex pattern stabilization in thin films resulting from shear thickening of active suspensions}

\author{Henning Reinken}
\email{henning.reinken@ovgu.de}
\affiliation{Institut f\"ur Physik, Otto-von-Guericke-Universität Magdeburg, Universitätsplatz 2, 39106 Magdeburg, Germany} 

\author{Andreas M. Menzel}
\email{a.menzel@ovgu.de}
\affiliation{Institut f\"ur Physik, Otto-von-Guericke-Universität Magdeburg, Universitätsplatz 2, 39106 Magdeburg, Germany}

\date{\today}

\begin{abstract}
The need for structuring on micrometer scales is abundant, for example, in view of phononic applications. We here outline a novel approach based on the phenomenon of active turbulence on the mesoscale. As we demonstrate, a shear-thickening carrier fluid of active microswimmers intrinsically stabilizes regular vortex patterns of otherwise turbulent active suspensions. 
The fluid self-organizes into a
periodically structured nonequilibrium state.
Introducing additional passive particles of intermediate size leads to regular spatial organization of these objects. Our approach opens a new path towards functionalization through patterning of thin films and membranes.

\end{abstract}

\maketitle

Structuring materials on micro- and submicrometer scales is of central importance to various types of prospective applications of functionalized components, such as shape-changing nematic elastomers~\cite{sawa2011shape,ware2015voxelated,gladman2016biomimetic,giudici2022multiple} or phononic metamaterials~\cite{hussein2014dynamics}.
The latter provide, for instance, acoustic bandgaps~\cite{baumgartl2007tailoring,caleap2014acoustically,oudich2023tailoring}.
This function relies on spatially organizing colloidal particles on scales responsive to acoustic excitations.
Generally, such structuring requires measures imposed from outside. We here introduce a strategy that facilitates intrinsic patterning through nonequilibrium effects. It relies on self-supported regular organization of vortices in an otherwise turbulent suspension of active microswimmers. Key to this mechanism is a shear-thickening carrier fluid. 

Suspensions of microswimmers are a subclass of active matter~\cite{marchetti2013hydrodynamics,bechinger2016active, doostmohammadi2018active,bar2020self,gompper20202020}. The interplay with the surrounding fluid determines both the swimming behavior of individual swimmers and their mutual interactions.
These nonequilibrium systems are amenable to structure formation.
For example, bacterial suspensions develop turbulent states, swirling, and vortex formation~\cite{dombrowski2004self,sokolov2007concentration,cisneros2007fluid,sokolov2012physical,reinken2018derivation,alert2021active} despite prevailing low-Reynolds-number conditions.
Being able to control or switch between different patterns is important for possible applications such as  microscale extraction of work~\cite{sokolov2010swimming,kaiser2014transport}, microfluidic mixing~\cite{kim2007controlled,suh2010review}, or cargo transport~\cite{trivedi2015bacterial,sokolov2015individual,schwarz2017hybrid}. Especially, regular mesoscale patterning is eminent when creating functionalized materials.
Previous experimental and theoretical studies have shown that external fields~\cite{reinken2019anisotropic} or geometrical constraints such as coupled flow chambers~\cite{wioland2013confinement,lushi14fluid,wioland2016ferromagnetic} or small obstacles~\cite{nishiguchi2018engineering,sone2019anomalous,reinken2020organizing,zhang2020oscillatory} can stabilize regular vortex patterns~\cite{nishiguchi2018engineering,reinken2020organizing,reinken2022ising}.
However, it is desirable to achieve such regular structure formation intrinsically, without the need of external intervention. 

Many previous considerations on microswimmers and their collective behavior assume the solvent to be Newtonian, although many biological fluids actually exhibit non-Newtonian rheology or viscoelasticity.
Few recent exceptions deal with the effects of viscoelasticity~\cite{teran2010viscoelastic,datt2018two,puljiz2019memory,yasuda2020reciprocal,li2021microswimming, eberhard2023reciprocal}. Non-Newtonian behavior, such as shear thickening and shear thinning, have been addressed~\cite{montenegro2013physics,qiu2014swimming,li2015undulatory,datt2015squirming,mathijssen2016upstream,van2022effect}, but
only a limited number of studies explores resulting collective dynamics~\cite{bozorgi2011effect,bozorgi2013role,bozorgi2014effects,hemingway2015active,hemingway2016viscoelastic,li2016collective,
plan2020active,liu2021viscoelastic}.
The impact of the non-Newtonian effect of shear thickening on the complex pattern formation in active fluids has not been explored so far. 

Here, we turn to this open question, based on previous descriptions of the dynamics in suspensions of active microswimmers~\cite{slomka2015generalized,slomka2017geometry,slomka2017spontaneous,slomka2018nature}.
Past investigations correctly predicted the main features of mesoscale turbulence~\cite{wensink2012meso}, a dynamic state of vortex formation on an intermediate length scale much larger than the single-swimmer scale~\cite{wensink2012meso,sokolov2012physical}.
We now incorporate non-Newtonian effects by a viscosity that increases with local shear rate.
Our results show that such shear thickening stabilizes regular structures, specifically centered rectangular lattice-like patterns consisting of elongated vortices.
Remarkably, geometrical constraints or other externally applied means of control are not necessary for this dynamic rotational symmetry breaking associated with anisotropic regular pattern formation.
Moreover, introducing passive particles larger than the active swimmers leads to their spatial organization and regular patterning according to the vortex patterns. In this way, we reveal a novel path towards structuring and functionalization of thin metamaterials.

\begin{figure*}
\includegraphics[width=0.999\linewidth]{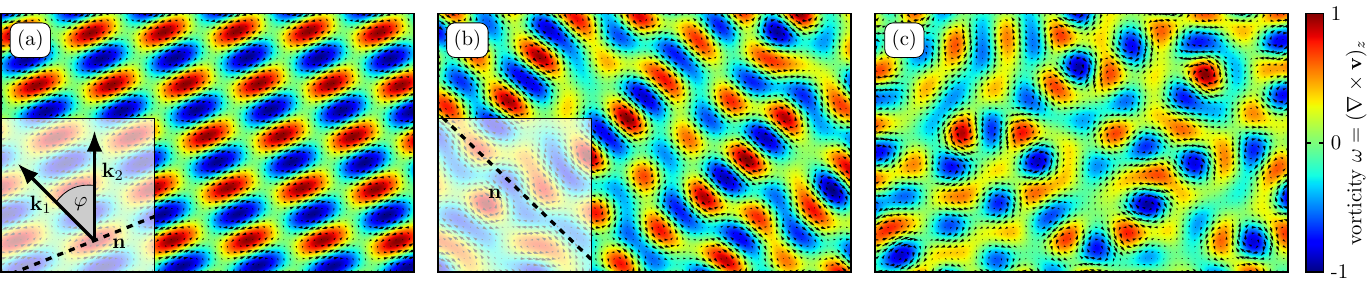}
\caption{\label{fig: snapshots shear thickening}Snapshots of the vorticity field $\omega(\mathbf{x},t)$ at $a = 1.1$ and $n = 3$ at arbitrarily chosen times $t$ for different values of the reference shear rate, (a) $\dot{\gamma}_2 = 0.1$, (b) $\dot{\gamma}_2 = 0.9$, and (c) $\dot{\gamma}_2 = 2.5$. 
Arrows denote the velocity field $\mathbf{v}$. 
Both quantities are rescaled for visualization purposes.
For very small $\dot{\gamma}_2$, the flow field settles into a stationary vortex lattice, where vortices are elongated along the director $\mathbf{n}$ shown as the dashed line.
The structure can be represented by two wavevectors forming an angle $\varphi = \pi/4$ (inset).
For larger $\dot{\gamma}_2$ the flow field becomes increasingly irregular. The size of the snapshots is $16\pi\times10\pi$.}
\end{figure*}

We describe the dynamics of the active suspension by a generalized, incompressible Navier--Stokes equation for the overall velocity field $\mathbf{v}(\mathbf{x},t)$ of the entire suspension,  
\begin{equation}
\label{eq: NSt}
\partial_t \mathbf{v} + \mathbf{v} \cdot \nabla \mathbf{v} = - \nabla \tilde{p} + \nabla \cdot \tilde{\boldsymbol{\sigma}}\, , \qquad \nabla \cdot \mathbf{v} = 0 \, ,
\end{equation}
rendering the approach Galilei invariant. Here, the constant density $\rho$ is absorbed into the pressure $p$ ($\tilde{p} = p/\rho$) and the stress tensor $\boldsymbol{\sigma}$ ($\tilde{\boldsymbol{\sigma}} = \boldsymbol{\sigma}/\rho$).
Following recent achievements to include the energy input by the microswimmers~\cite{slomka2015generalized,slomka2017geometry,slomka2017spontaneous}, we expand the stress tensor in gradients of the deformation rate $\boldsymbol{\Sigma} = [(\nabla \mathbf{v}) + (\nabla \mathbf{v})^\top]/2$,
\begin{equation}
\label{eq: stress expansion}
\tilde{\boldsymbol{\sigma}} = ( \Gamma_0 + \Gamma_2 \nabla^2 + \Gamma_4 \nabla^4 ) \big[ (\nabla \mathbf{v}) + (\nabla \mathbf{v})^\top \big]\, ,
\end{equation}
where $^\top$ marks the transpose.
Often, the active stress %
includes orientational order parameter fields, which are governed by additional dynamic equations~\cite{reinken2018derivation,doostmohammadi2018active}.
In our case, the orientational order parameters are ``slaved'' to the suspension velocity~\cite{slomka2017spontaneous}.
In principle, one can %
include terms $\propto \mathbf{v}\mathbf{v} - |\mathbf{v}|^2\mathbf{I}/d$, where $\mathbf{I}$ is the unit tensor and $d$ spatial dimensionality.
However, these terms would only lead to a rescaling of the nonlinear advection term and an additional contribution to the pressure~\cite{slomka2015generalized}.
Equation~(\ref{eq: stress expansion}) captures essential experimental observations on active suspensions, for example, length scale selection and emergence of turbulent vortex patterns~\cite{slomka2015generalized}. 
Very good agreement with both bacterial microswimmers and ATP-driven microtubular networks has been demonstrated~\cite{slomka2017spontaneous}.
Recent work on \textit{passive} suspensions showed that higher-order gradients of the suspension-averaged velocity can emerge in the effective stress tensor within a rigorous derivation~\cite{wolgemuth2023continuum}, which further supports our approach.

Positive values of the coefficients $\Gamma_0$ and $\Gamma_4$ ensure asymptotic stability at long and short wavelengths, respectively~\cite{slomka2015generalized}.
In the passive case, $\Gamma_4 = 0$, $\Gamma_2 = 0$, and $\Gamma_0= \nu$, introducing the kinematic viscosity $\nu$. 
In the active case, we adopt $\Gamma_0 =\nu$, while  $\Gamma_2$ can show either sign. 
For $\Gamma_2 < 0$, there is no active energy input and the quiescent state $\mathbf{v}(\mathbf{x},t) = \mathbf{0}$ is stable. Instead, for $\Gamma_2 > 0$, active stresses set in, which can excite intermediate wavelengths~\cite{slomka2015generalized}.
Thus, $\Gamma_2$ characterizes the strength of activity.
When increasing $\Gamma_2 > \sqrt{4 \nu \Gamma_4}$, linear stability analysis yields a finite-wavelength instability of critical wavenumber $k_\mathrm{c} = \sqrt{\Gamma_2/(2\Gamma_4)}$.
Further increasing $\Gamma_2$, a band of unstable modes emerges, indicating wavenumbers at which activity pumps energy into the system.
Similarly to driven Navier--Stokes fluids~\cite{davidson2015turbulence}, the nonlinear advection term $\mathbf{v}\cdot\nabla\mathbf{v}$ relates to turbulence and energy transport between wavenumbers. 
However, driving in our case is internal, due to the active energy input by the microswimmers.
Resulting balances of active energy input and dissipation lead to statistically stationary states, in line with main features of experimental observations on bacterial suspensions~\cite{slomka2015generalized}.
Recent studies employing a similar description suggest a transition between different spectral scaling regimes in active turbulence upon an increase of activity~\cite{mukherjee2023intermittency}.
In this context, we here focus on the \textit{mildly} active regime, see Ref.~\cite{mukherjee2023intermittency}, where the statistics of velocity increments follows a Gaussian distribution.

Shear thickening of the carrier liquid is described by a viscosity increasing with local shear rates $\dot{\gamma}(\mathbf{x})$.
We consider a so-called power-law fluid~\cite{waele1923viscometry,ostwald1925ueber,irgens2014rheology} of constant zero-shear viscosity $\nu_0$,
\begin{equation}
\label{eq: shear thickening viscosity}
\nu(\mathbf{x}) = \nu_0 + \nu_0 \bigg(\frac{\dot{\gamma}(\mathbf{x})}{\dot{\gamma}_2}\bigg)^{n-1}\, ,
\end{equation}
$\dot{\gamma}(\mathbf{x}) = \sqrt{2\, \boldsymbol{\Sigma}(\mathbf{x}):\boldsymbol{\Sigma}(\mathbf{x})}.$ %
The exponent $n$ determines how the viscosity $\nu(\mathbf{x})$ increases with local shear rate $\dot{\gamma}(\mathbf{x})$.  $\dot{\gamma}_2$ is a reference shear rate, indicating when the viscosity reaches $2\nu_0$. 

We now rescale lengths by $k_\mathrm{c}^{-1}$, times by $(k_\mathrm{c}^2 \nu_0)^{-1}$, and thus velocities by $k_\mathrm{c} \nu_0$.
As a result, Eq.~(\ref{eq: NSt}) becomes
\begin{equation}
\label{eq: generalized NSt rescaled}
\partial_t \mathbf{v} + \mathbf{v} \cdot \nabla \mathbf{v} = - \nabla \tilde{p} + \nabla \cdot (2 \nu \boldsymbol{\Sigma}) + a (2\nabla^4 \mathbf{v} + \nabla^6 \mathbf{v})\, ,
\end{equation}
where $\nu = 1 + (\sqrt{2\, \boldsymbol{\Sigma}:\boldsymbol{\Sigma}}/\dot{\gamma}_2)^{n-1}$. 
In our incompressible system,  $\tilde{p}$  merely acts as a Langrange multiplier ensuring $\nabla \cdot \mathbf{v} = 0$.
The parameter $a = \Gamma_2^2/(4 \nu_0 \Gamma_4)$ sets the strength of active energy input relative to the zero-shear viscosity $\nu_0$.
For $a < 1$, the isotropic, quiescent state $\mathbf{v}(\mathbf{x},t) = \mathbf{0}$ is stable, implying that the active energy input does not suffice to overcome viscous dissipation.
Thus, pattern formation is not observed.
Conversely, for $a > 1$, the system forms flow patterns characterized by a specific length scale set by the fastest-growing mode $k_\mathrm{m}$.
Close to the transition, $k_\mathrm{m}=k_\mathrm{c} = 1$, yielding a length scale of $\Lambda_\mathrm{c} = 2\pi/k_\mathrm{c}$.
For $a = 0$, we recover the passive Navier--Stokes equation for non-Newtonian, shear-thickening behavior.

\begin{figure*}
\includegraphics[width=0.999\linewidth]{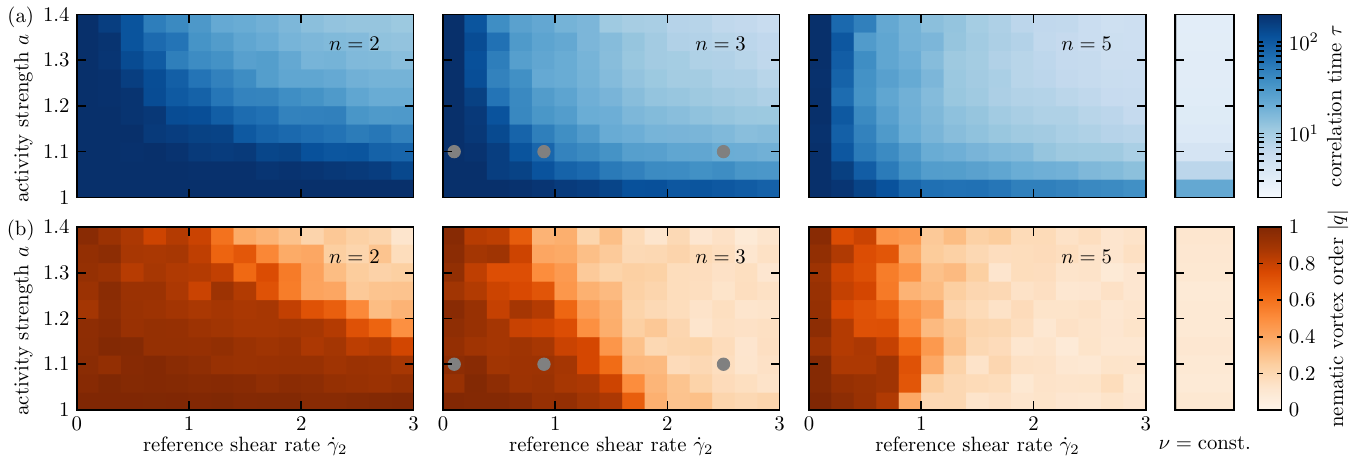}
\caption{\label{fig: state diagrams}Characteristic quantities as a function of $a$ and $\dot{\gamma}_2$ for different values of $n$. 
(a) Correlation time $\tau$ on a log-scale.
(b) Nematic order parameter $|q|$ for the elongated vortices.
High nematic order for small values of $\dot{\gamma}_2$ signifies a globally ordered vortex structure with broken rotational symmetry.
The gray circles in the plots for $n=3$ indicate the parameter values where the snapshots shown in Fig.~\ref{fig: snapshots shear thickening} are taken.
The column on the right-hand side includes the Newtonian case ($\nu = 1$) for comparison.}
\end{figure*}

We employ a pseudo-spectral method to solve Eq.~(\ref{eq: generalized NSt rescaled}) in a two-dimensional system with periodic boundary conditions starting from random initial values, for details see the Supplemental Material~\cite{suppl}, which includes Refs.~\cite{dunkel2013fluid,maxey1983equation,brandt2022particle,haller2008inertial,balachandar2010turbulent,eaton1994preferential}. %
The system size is set to $48\pi\times 48\pi$, much larger than the critical length scale $\Lambda_\mathrm{c} = 2\pi$.
Varying the values of activity $a$, power-law exponent $n$, and reference shear rate $\dot{\gamma}_2$, we investigate the emerging patterns.
Examples are illustrated in Fig.~\ref{fig: snapshots shear thickening}, where snapshots of the vorticity field  $\omega = (\nabla \times \mathbf{v})_z$ are shown for $a = 1.1$, $n = 3$, and varying $\dot{\gamma}_2$.

For high reference shear rate $\dot{\gamma}_2$, that is, closer to Newtonian behavior, the system develops a turbulent state, similar to the case without shear thickening~\cite{slomka2015generalized,slomka2017geometry,slomka2017spontaneous}, see Fig.~\ref{fig: snapshots shear thickening}(c).
However, when $\dot{\gamma}_2$ is low, a rather regular state emerges, which can be characterized as a centered rectangular lattice of vortices.
Figure~\ref{fig: snapshots shear thickening}(a) shows the vorticity field in this state.
The vortices become elongated along a common axis, displaying an aspect ratio of about $3$.
For intermediate values of $\dot{\gamma}_2$, the system develops a state of still clearly visible anisotropy of the vortices, see Fig.~\ref{fig: snapshots shear thickening}(b), yet not stationary.
Instead, numerous defects and dynamic reorganization occur.

To further characterize the observed spatiotemporal patterns, we first calculate the correlation time $\tau$~\cite{suppl}, which quantifies how quickly the velocity field reorganizes.
Figure~\ref{fig: state diagrams}(a) shows $\tau$ in $a$-$\dot{\gamma}_2$-space for different values of $n$.
Consistently with our previous observations, we find that $\tau$ is small for large values of both $a$ and $\dot{\gamma}_2$, indicating a turbulent state.
Decreasing either $a$ or $\dot{\gamma}_2$ increases $\tau$, and the dynamics becomes slower until the emerging patterns settle into a stationary state for very small values of $a$ and $\dot{\gamma}_2$.
Since the system does not rearrange anymore, $\tau$ diverges. 
Besides, $\tau$ tends to decrease for increasing $n$, so that smaller $n$ stabilize the regular elongated vortex structure.
For comparison, $\tau$ as a function of $a$ is shown on the right-hand side of Fig.~\ref{fig: state diagrams}(a) for a Newtonian suspension without shear thickening ($\nu = 1$).
In this case, larger values of $a$ lead to an increased input of active energy into the system and, thus, to a more turbulent state.

In Fig.~\ref{fig: snapshots shear thickening}(a) and (b), the elongated vortices on average align along a common axis. 
Thus, we determine the global nematic order parameter $q$ for orientational order of elongated vortices~\cite{suppl}.
$|q|=0$ for a flow field of uniformly distributed vortex orientations, whereas $|q|=1$ for completely ordered systems, such as in Fig.~\ref{fig: snapshots shear thickening}(a).
Figure~\ref{fig: state diagrams}(b) shows $|q|$ for different values of $n$. 
For smaller reference shear rate $\dot{\gamma}_2$, we indeed find that the elongated vortices are aligned along a common axis $\mathbf{n}$, see Fig.~\ref{fig: snapshots shear thickening}(a,b).
Thus, shear thickening does not only lead to local vortex elongation, but also to spontaneous overall rotational symmetry breaking.
The system exhibits global nematic order even for values of $\dot{\gamma}_2$ associated with defects and reorganization as in Fig.~\ref{fig: snapshots shear thickening}(b).
For developed turbulence, $|q|$ approaches zero, see the right-hand side of Fig.~\ref{fig: state diagrams}(b).
Again, we find that increasing $n$ reduces stabilization of the elongated vortex pattern and thus suppresses $|q|$. 
Moreover, the impact of activity $a$ on stabilization seems to diminish.

\begin{figure}
\includegraphics[width=0.999\linewidth]{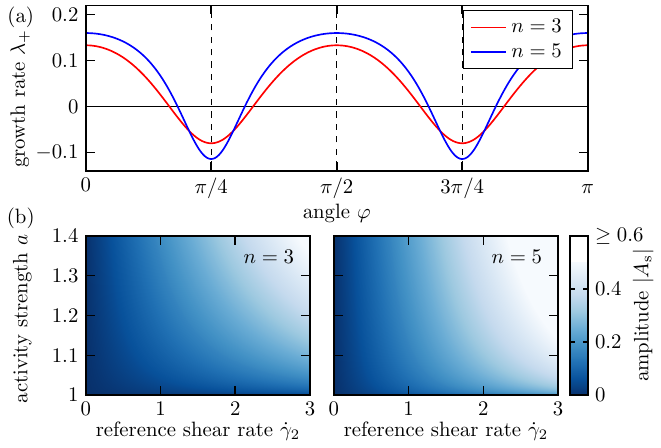}
\caption{\label{fig: growth rate and amplitude}Results obtained from the amplitude equations. 
(a) Maximum growth rate of perturbations with respect to the stationary solution for $n=3$ [Eq.~(\ref{eq: eigenvalues growth rate n=3})] and $n=5$~\cite{suppl} as a function of the angle $\varphi$ between the two modes representing the vortex pattern.
(b) Amplitude $|A_\mathrm{s}|$ of the stationary solution at $\varphi = \pi/4$ as a function of $a$ and $\dot{\gamma}_2$ for $n=3$ [Eq.~(\ref{eq: stationary amplitude n=3})] and for $n=5$~\cite{suppl}.}
\end{figure}

Conceptually, the shear-thickening properties of the suspension provide a saturation mechanism for the growing vortex patterns that does not rely on the turbulent energy transfer to larger scales and subsequent dissipation.
As a result, regular structures are stabilized.
To shed more light on these effects, we determine associated amplitude equations. 
The stationary, centered rectangular lattice close to its emergence forms an orthorhombic state that is represented by two modes of complex amplitudes $A_1$ and $A_2$ and wavevectors $\mathbf{k}_1$ and $\mathbf{k}_2$~\cite{cross2009pattern}.
As inferred from the linear stability analysis outlined above, the isotropic state becomes unstable to the growth of perturbations of wavevectors $|\mathbf{k}| = k_\mathrm{c} = 1$ when activity is increased above $a = 1$.
Close to this critical point, we assume the length scale $\Lambda_\mathrm{c} = 2\pi/k_\mathrm{c}$ to dominate the emerging patterns.
We write the wavevectors of the modes $\mathbf{k}_i$ ($i=1,2$) as $\mathbf{k}_i = k_\mathrm{c}(\cos\varphi_i,\sin\varphi_i)$. %
Without loss of generality, we set one of the angles to zero, $\varphi_1 = 0$.
Then $\varphi_2 = \varphi$ determines the relative angle between the wavevectors.
Thus, vortex structures as in Fig.~\ref{fig: snapshots shear thickening}(a) are parameterized by
\begin{align}
\label{eq: two modes}
v_x &= A_2 \sin(\varphi) \, e^{i[\cos(\varphi)x + \sin(\varphi)y]} + \mathrm{c.c.} \, ,\\
v_y &= - A_1 e^{ix} - A_2 \cos(\varphi) \, e^{i[\cos(\varphi)x + \sin(\varphi)y]} + \mathrm{c.c.} \, , \nonumber
\end{align}
satisfying incompressibility, where $\mathrm{c.c.}$ denotes complex conjugates.

Inserting Eq.~(\ref{eq: two modes}) into Eq.~(\ref{eq: generalized NSt rescaled}), we restrict ourselves to values $n = 3$ and $5$ to make analytical progress, see also Ref.~\cite{suppl}. %
Using symbolic computation software~\cite{sympy}
and collecting terms $\sim e^{ix}$ and $\sim e^{i[\cos(\varphi)x + \sin(\varphi)y]}$,  %
we find for $n = 3$ \cite{suppl}
\begin{equation}
\label{eq: amplitude equations n=3}
\begin{aligned}
\frac{\partial A_1}{\partial t} = (a - 1)A_1 - \frac{A_1}{\dot{\gamma}_2^2}\big\{3|A_1|^2 + 2|A_2|^2\big[2 + \cos(4 \varphi)\big] \big\}\, ,\\
\frac{\partial A_2}{\partial t} = (a - 1)A_2 - \frac{A_2}{\dot{\gamma}_2^2}\big\{3|A_2|^2 + 2|A_1|^2\big[2 + \cos(4 \varphi)\big] \big\}\, . 
\end{aligned}
\end{equation}
Assuming $|A_1| = |A_2| = |A_\mathrm{s}|$, the nontrivial amplitude of the  stationary solution becomes
\begin{equation}
\label{eq: stationary amplitude n=3}
|A_\mathrm{s}| = \dot{\gamma}_2\bigg[\frac{a - 1}{7 + 2 \cos(4\varphi)}\bigg]^{\frac{1}{2}}\, .
\end{equation}
A linear stability analysis yields the resulting decay and/or growth rates
\begin{equation}
\label{eq: eigenvalues growth rate n=3}
\lambda_- = - 2 (a-1)\, , \quad \lambda_+ = \frac{2(a -1)[1 + 2 \cos(4\varphi)]}{7 + 2\cos(4\varphi)}\, ,
\end{equation}
where only $\lambda_+$ can become positive for $a > 1$ and, thus, determines stability.

Whether the stationary solution is stable is determined by the angle $\varphi$, see Fig.~\ref{fig: growth rate and amplitude}(a).  
In particular, perpendicular configurations of $\varphi = \pi/2$ are unstable ($\lambda_+ > 0$), whereas configurations of $\varphi = \pi/4$ are stable ($\lambda_+ < 0$).
These analytical results confirm our numerical observations and explain the geometry of the vortex lattice. 
Therefore, shear thickening %
leads to skewed lattices of $\varphi \neq \pi/2$ and vortex elongation along a common axis $\mathbf{n}$.
In Fig.~\ref{fig: snapshots shear thickening}(a), $\mathbf{n}$ is inclined  by an angle of $3\pi/8$ from each wavevector.
The case of  $n=5$ \cite{suppl} leads to similar results, %
yet with a slightly narrower region of stability, %
 see Fig.~\ref{fig: growth rate and amplitude}(a).

Due to the nonlinear nature of advection, destabilization becomes more important with increasing amplitude of the emerging patterns.
To explore this point, we plot the stationary amplitude according to Eq.~(\ref{eq: stationary amplitude n=3}) for $n=3$ and %
for $n=5$ \cite{suppl} in Fig.~\ref{fig: growth rate and amplitude}(b).
It grows with increasing activity $a$ and reference shear rate $\dot{\gamma}_2$.
This is caused by pattern saturation being mediated via shear-thickening effects and thus becoming stronger when these set in earlier, that is, for smaller $\dot{\gamma}_2$.
Comparing Fig.~\ref{fig: growth rate and amplitude}(b) with the correlation times $\tau$ displayed in Fig.~\ref{fig: state diagrams}(a) adds to this point.
Increasing $a$ or $\dot{\gamma}_2$ leads to faster dynamics of the flow field and thus to a more turbulent state.
The dependence of $A_\mathrm{s}$ on the reference shear rate $\dot{\gamma}_2$ is linear in Eq.~(\ref{eq: stationary amplitude n=3}). %
This implies substantial impact of variations in $\dot{\gamma}_2$ on the emerging spatiotemporal structures, in line with the results for the correlation time $\tau$ and the degree of nematic order $|q|$ shown in Fig.~\ref{fig: state diagrams} for different values of $n$.

\begin{figure}
\includegraphics[width=0.999\linewidth]{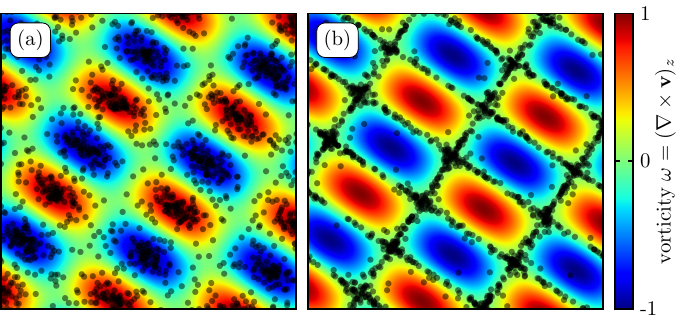}
\caption{\label{fig: passive particles}
Spatial organization of passive particles (dark dots) of diameter $d=0.06\Lambda_\mathrm{c}$ at $\mathrm{Stk} = 0.033$ in an active shear-thickening carrier fluid at time $t = 16000$ after starting from random initial distributions. (a) Lighter particles (here half the fluid density, $R = 1$) cluster in vortex centers, whereas (b) heavier particles (here twice the fluid density, $R = 2/5$) accumulate between vortices. Snapshot size is $6\pi\times 6\pi$.}
\end{figure}

As an immediate perspective, the emerging regular vortex patterns facilitate the spatial organization of objects within the system.
To demonstrate this effect, we consider the dynamics of passively advected particles of intermediate size and perform additional simulations using a simplified form of the Maxey--Riley equation~\cite{maxey1983equation,haller2008inertial} coupled to the flow field obtained via Eq.~(\ref{eq: generalized NSt rescaled}), see Ref.~\cite{suppl} for details.
As is known from particle-laden flows~\cite{maxey1983equation,haller2008inertial,balachandar2010turbulent,brandt2022particle}, passive objects tend to accumulate in certain areas of the flow if their density, $\rho_\mathrm{p}$, is different from that of the carrier fluid, $\rho_\mathrm{s}$.
Lighter objects cluster within vortices, whereas heavier objects are ejected from them~\cite{eaton1994preferential}.
The density ratio is characterized by the factor $R = 2\rho_\mathrm{s}/(\rho_\mathrm{s} + 2\rho_\mathrm{p})$.
We further introduce the Stokes number $\mathrm{Stk}$, which measures the characteristic time scale of the particle dynamics relative to that of the flow.
Under present scaling, $\mathrm{Stk}$ is defined via $\mathrm{Stk} = 8\pi^2 d^2/(9\Lambda_\mathrm{c}^2)$, which grows quadratically with the particle diameter and vanishes for point particles~\cite{suppl}.
These effects in combination with the self-supported regular pattern formation of our active shear-thickening carrier fluid can be used to spatially organize objects into regular periodic structures, see Fig.~\ref{fig: passive particles} for snapshots from the numerical simulations at $\mathrm{Stk} = 0.033$ and different density ratios.
Thus, intrinsic pattern formation in shear-thickening active suspensions opens a new strategy of generating sheets of functionalized metamaterials based on regular positional structuring of embedded objects on the microscale~\cite{baumgartl2007tailoring,oudich2023tailoring}.

To summarize, we reveal that shear thickening is able to reorganize the flow field in active suspensions and intrinsically stabilizes regular vortex patterns in an otherwise turbulent state.
Elongation arises for the vortices along a common axis. %
In contrast to other related observations on suspensions of active microswimmers, the patterns here are intrinsically stabilized by the shear-thickening carrier liquid and do not require external stabilization via geometrical constraints, such as arrangements of small obstacles~\cite{nishiguchi2018engineering,sone2019anomalous,reinken2020organizing,zhang2020oscillatory,reinken2022ising}, systems of coupled flow chambers~\cite{wioland2013confinement,lushi14fluid,wioland2016ferromagnetic} or substrate friction~\cite{doostmohammadi2016stabilization}.
The effect can therefore be employed to intrinsically structure functionalized components on the micrometer scale. 
Changing the properties of the microswimmers, for instance, self-swimming speed
or body size, allows to tune the intrinsic selection of length scales~\cite{sokolov2012physical} and thus the lattice constant of the stabilized
patterns.

\begin{acknowledgments}

We thank the Deutsche Forschungsgemeinschaft (DFG, German Research Foundation) for support of this work through the Research Grant ME 3571/5-1. A.M.M.\ acknowledges support by the Deutsche Forschungsgemeinschaft through the Heisenberg Grant ME 3571/4-1. 

\end{acknowledgments}

\end{document}


\title{Vortex pattern stabilization in thin films resulting from shear thickening of active suspensions\\[1\baselineskip]
\textit{Supplemental Material}}

\author{Henning Reinken}
\email{henning.reinken@ovgu.de}
\affiliation{Institut f\"ur Physik, Otto-von-Guericke-Universität Magdeburg, Universitätsplatz 2, 39106 Magdeburg, Germany} 

\author{Andreas M. Menzel}
\email{a.menzel@ovgu.de}
\affiliation{Institut f\"ur Physik, Otto-von-Guericke-Universität Magdeburg, Universitätsplatz 2, 39106 Magdeburg, Germany}

\date{\today}

\maketitle

\section{Numerical methods}
\label{app: numerical methods}

We solve Eq.~(4) in a two-dimensional system with periodic boundary conditions employing a pseudo-spectral method to increase accuracy and speed of the evaluation of spatial derivatives.
The spatial resolution is set to $400\times 400$ grid points.
Time integration is performed via a fourth-order Runge-Kutta method, which is combined with an operator splitting technique
where the linear and nonlinear parts are treated consecutively.
Due to the incompressibility condition $\nabla \cdot \mathbf{v} = 0$, the pressure $p$ effectively becomes a Lagrange multiplier that can be used to ensure that $\mathbf{v}(\mathbf{x},t)$ stays divergence-free. 
Every calculation is started with a quiescent velocity field, i.e., $\mathbf{v}(\mathbf{x},t) = \mathbf{0}$, with the addition of a small random perturbation at every grid point taken from a uniform distribution over $[-0.01,0.01]$.
The system size is set to $48\pi\times48\pi$, which allows for more than $500$ vortices in the system, at least in the case of highly ordered states.

Before starting the analysis of the emerging structures, we always wait for $10,000$ units of rescaled time to ensure that the system has developed a statistically stationary state.
The characteristic quantities shown in the diagrams in Fig.~2 in the main text are obtained by repeating the calculations for every pair of parameter values $a$ and $\dot{\gamma}_2$ twelve times starting from different realizations of the random perturbations to the initial state.
The calculated quantities are then averaged over these realizations.

To analyze the emerging structures, we first calculate the temporal correlation function of the flow fields,
\begin{equation}
\label{eq: Eulerian temporal correlation}
C(\Delta t) = v^{-2}\langle \mathbf{v}(\mathbf{x},t) \cdot \mathbf{v}(\mathbf{x},t + \Delta t)\rangle \, .
\end{equation}
Here, $\Delta t$ denotes a time a lag, $v$ is the mean speed averaged over the whole system, and the average $\langle \dots \rangle$ is performed over both time $t$ and space $\mathbf{x}$.
The correlation time $\tau$, which characterizes how quickly the velocity field reorganizes, is obtained via 
\begin{equation}
\label{eq: Eulerian correlation time}
\tau = \int_0^\infty C(\Delta t) \; d \Delta t \, .
\end{equation}

Before calculating the nematic vortex order parameter $|q|$, we must first determine the orientations of the vortices from a snapshot of the vorticity field $\omega(\mathbf{x},t)$ at a certain time $t$.
To this end, we identify vortices by searching for local maxima of the absolute vorticity $|\omega|(\mathbf{x},t)$.
Only fully developed vortices are taken into account. We identify a local maximum as the center point of a vortex $j$ when the vorticity $|\omega|_j$ at that point is more than double the mean absolute vorticity $\langle |\omega| \rangle$ (averaged over the whole system).
Then, as a function of the direction set by the angle $\varphi$ and measured from the center of the vortex, we determine the distance $\ell_j(\varphi)$ where $|\omega|$ has decayed to half of its value $|\omega|_j$.
We calculate the ratio
\begin{equation}
\label{eq: ratio axes vortex}
r_j(\varphi) = \frac{\ell_j(\varphi)}{\ell_j(\varphi + \pi/2)}\, ,
\end{equation}
which compares the extension of the vortex along mutually perpendicular directions.
The angle of maximum $r_j(\varphi)$ is identified as the orientation $\theta_j$ of the long axis of the vortex $j$.
We finally determine the global nematic order parameter $q$ for orientational order of elongated vortices as the complex number
\begin{equation}
\label{eq: alignment order parameter}
q = \frac{1}{N} \sum_{j=1}^N e^{2i\theta_j} \, .
\end{equation}
Here, $N$ is the total number of vortices.
The orientation of the nematic director is determined by the angle $\varphi_q = \arg(q)/2$ via $\mathbf{n} = (\cos\varphi_q, \sin\varphi_q)$.

The evolution equations for the positions $\mathbf{X}_i$ and velocities $\mathbf{U}_i$ of the passive particles within the flow field (see section~\ref{sec: particles} of this supplemental) are propagated in time using a fourth-order Runge-Kutta scheme.
Initially, particles are distributed uniformly within the simulation box and their initial velocities are taken as the values of the velocity field $\mathbf{v}$ at their positions.

\section{Derivation of amplitude equations}
\label{app: derivation of amplitude equations}

In order to derive amplitude equations, we insert the representation of the vortex lattices,  
Eqs.~(8), into Eq.~(4). Then, we collect terms containing identical modes as in the vortex lattice representation. 
We here provide details of this procedure in the case of $n=3$.

The ansatz in Eqs.~(8) fulfills the incompressibility condition, $\nabla \cdot \mathbf{v} = 0$.
However, the nonlinear terms can lead to non-zero divergence, which in turn needs to be compensated by the pressure $p$.
We expand the pressure to reflect the representation in Eqs.~(8) according to
\begin{align}
\label{eq: pressure expansion}
p = p_1 e^{ix} + p_2 e^{i[\cos(\varphi)x + \sin(\varphi)y]} + \mathrm{c.c.} \, .
\end{align}
Inserting Eqs.~(8) and (\ref{eq: pressure expansion})  into Eq.~(4) and calculating the divergence of the right-hand side, which must vanish, we find expressions for the  amplitudes $p_1$ and $p_2$.
In particular, collecting terms %
$\sim \exp\{ix\}$ yields
\begin{align}
\label{eq: pressure mode 1}
p_1 = \frac{2i}{\dot{\gamma}_2^2} |A_2|^2 A_1 \sin(4\varphi) \, , 
\end{align}
whereas %
from terms $\sim \exp\{i[\cos(\varphi)x + \sin(\varphi)y]\}$ we find
\begin{align}
\label{eq: pressure mode 2}
p_2 = {}-\frac{2i}{\dot{\gamma}_2^2} |A_1|^2 A_2 \sin(4\varphi) \, .
\end{align}

Having determined the relation between  amplitudes $p_1$, $p_2$, $A_1$, and $A_2$, we turn back to Eq.~(4) after insertion of Eqs.~(8) and (\ref{eq: pressure expansion}).
We derive the amplitude equations by collecting terms $\sim \exp\{ix\}$ and $\sim \exp\{i[\cos(\varphi)x + \sin(\varphi)y]\}$.
This procedure yields four equations,
\begin{align}
\label{eq: amplitude equations n=3 preliminary x 1}
0 = &- \frac{2}{\dot{\gamma}_2} |A_2|^2 A_1 \sin(4\varphi) - i p_1 \, , \\
\label{eq: amplitude equations n=3 preliminary y 1}
\frac{\partial A_1}{\partial t} = &(a - 1)A_1  - \frac{A_1}{\dot{\gamma}_2^2}\big\{3|A_1|^2 + 2|A_2|^2\big[2 + \cos(4 \varphi)\big] \big\}\, , \\
\label{eq: amplitude equations n=3 preliminary x 2}
\frac{\partial A_2}{\partial t} = &(a - 1)A_2 - \frac{i p_2}{\tan(\varphi)} -\frac{A_2}{\dot{\gamma}_2^2}\big\{ 3|A_2|^2+ 8 |A_1|^2 \sin^2(\varphi) - 2 |A_1|^2 \big\} \, , \\
\label{eq: amplitude equations n=3 preliminary y 2}
\frac{\partial A_2}{\partial t} = &(a - 1)A_2 + i p_2\tan(\varphi) -\frac{A_2}{\dot{\gamma}_2^2}\big\{ 3|A_2|^2 - 8 |A_1|^2 \sin^2(\varphi) + 6 |A_1|^2 \big\} \, .
\end{align}
Equation~(\ref{eq: amplitude equations n=3 preliminary x 1}) is satisfied by the condition for $p_1$, see Eq.~(\ref{eq: pressure mode 1}).
Equation~(\ref{eq: amplitude equations n=3 preliminary y 1}), which is independent of $p_1$, is the amplitude equation already included in the main text for $A_1$, see Eqs.~(9).
Inserting the expression for $p_2$, Eq.~(\ref{eq: pressure mode 2}), we find that both Eqs.~(\ref{eq: amplitude equations n=3 preliminary x 2}) and (\ref{eq: amplitude equations n=3 preliminary y 2}) yield the same result, which is the amplitude equation for $A_2$, see Eqs.~(9) as well.

For completeness, we note that evolution equations for $A_1^\ast$ and $A_2^\ast$ (the complex conjugates of $A_1$ and $A_2$) are derived using the same procedure but collecting terms %
$\sim\exp\{-ix\}$ and %
$\sim\exp\{-i[\cos(\varphi)x + \sin(\varphi)y]\}$, respectively.
In this way, we find that the evolution of $A_1^\ast$ and $A_2^\ast$ is governed by the same dynamics as for $A_1$ and $A_2$.
All coefficients in the derived amplitude equations are real.
Thus, we conclude that the phases of the two modes can be considered as free parameters that simply lead to a shift of the whole lattice-like structure.

We proceed with the derivation of amplitude equations for $n=5$. 
Applying the same procedure as outlined for $n=3$ yields
\begin{equation}
\label{eq: amplitude equations n=5}
\begin{aligned}
\frac{\partial A_1}{\partial t} = (a &- 1)A_1 - \frac{A_1}{\dot{\gamma}_2^4}\big\{10A_1^4 +(12A_1^2 A_2^2 + 6 A_2^4)\big[3 + 2\cos(4 \varphi)\big] \big\}\, ,\\
\frac{\partial A_2}{\partial t} = (a &- 1)A_2 - \frac{A_2}{\dot{\gamma}_2^4}\big\{10A_2^4 +(12A_1^2 A_2^2 + 6 A_1^4)\big[3 + 2\cos(4 \varphi)\big] \big\}\, ,
\end{aligned}
\end{equation}
which are then analyzed analogously.
First, the amplitude of the nontrivial, stationary solution is
\begin{equation}
\label{eq: stationary amplitude n=5}
|A_\mathrm{s}| = \dot{\gamma}_2\bigg[ \frac{(a - 1)}{64 + 36 \cos(4\varphi)}\bigg]^{\frac{1}{4}}\, .
\end{equation}
Again, performing a linear stability analysis, we find the decay and/or growth rate of perturbations to this solution as
\begin{equation}
\label{eq: eigenvalues growth rate n=5}
\lambda_- = - 4 (a-1)\, , \quad \lambda_+ = \frac{4(a -1)[2 + 3 \cos(4\varphi)]}{16 + 9\cos(4\varphi)}\, .
\end{equation}
The form of the eigenvalues $\lambda_\pm$ is structurally very similar to the one in Eq.~(11).
Again we find that the vortex pattern is only stable in a configuration with an angle between the two wavevectors  emerging at threshold close to $\varphi = \pi/4$.

In the derivation of the amplitude equations, the integer values $n=3$ and $n=5$ favor the analytical procedure. This is because, for odd integer values of $n$, the exponent in the nonlinear term $\propto (2\mathbf{\Sigma}:\mathbf{\Sigma})^{(n-1)/2}$ in the viscosity [see Eq.~(3) in the main text] becomes an integer. When inserting the ansatz for the regular patterns [see Eq.~(5) in the main text] into the formula for the viscosity, we obtain a sum of harmonics via a multinomial expansion. Specifically, the case of $n=3$ is particularly straightforward. Here we find
\begin{equation}
\begin{aligned}
(2\mathbf{\Sigma}:\mathbf{\Sigma})^{\frac{3-1}{2}} = 2\mathbf{\Sigma}:\mathbf{\Sigma} = &-A_1^2 e^{2ix} + 2 |A_1|^2 - A_1^{\ast 2}  e^{-2ix} -A_2^2 e^{2i[\cos(\varphi)x + \sin(\varphi)y]}  + 2 |A_2|^2 - A_2^{\ast 2} e^{-2i[\cos(\varphi)x + \sin(\varphi)y]} \\
&-2 A_1 A_2 \cos(2\varphi) e^{i [\cos(\varphi)x + x + \sin(\varphi)y]} +2 A_1 A_2^\ast \cos(2\varphi) e^{i [-\cos(\varphi)x + x - \sin(\varphi)y]} \\
&+2 A_1^\ast A_2 \cos(2\varphi)e^{i [\cos(\varphi)x - x + \sin(\varphi)y]} -2 A_1^\ast A_2^\ast \cos(2\varphi) e^{-i [\cos(\varphi)x + x + \sin(\varphi)y]}\, .
\end{aligned}
\end{equation}
Analytically, the procedure to derive the amplitude equation relies on expressing all terms as sums of harmonics. Matching the prefactors of these harmonics yields the amplitude equations to lowest order. Therefore, the procedure is analytically tractable for uneven integer values, especially in the simplest cases of $n=3$ and $n=5$ (and thus the procedure for even integer values of $n$, or noninteger values, becomes significantly more challenging).

\section{Transport of passive particles}
\label{sec: particles}

In order to explore how the vortex patterns impact the spatial organization of objects within the flow, we investigate the dynamics of $N$ passive spherical objects that are transported by the flow field.
Here, we focus on objects that are notably smaller than the characteristic length of the vortex patterns and, thus, do not significantly influence the flow field. At the same time, they are larger than the typical size of the active objects suspended in the shear-thickening fluid. Therefore, the active carrier fluid can still be described by a continuum approach. 

Although we are dealing with small systems on the micron scale, the bare diffusion of the suspended objects can be disregarded in our case.
To demonstrate the validity of this assumption, we estimate the Peclet number, which quantifies the ratio of advective to diffusive transport.
The Peclet number is determined via $\mathrm{Pe} = \Lambda_\mathrm{a} v_a / D_0$, where the length scale of the advecting flow and its average velocity are denoted by $\Lambda_\mathrm{a}$ and $v_\mathrm{a}$ and the bare diffusivity is denoted by $D_0$.
First, $D_0$ can be estimated at low Reynolds numbers via the Stokes--Einstein relation. Thus, $D_0 = k_\mathrm{B} T/(3\pi \eta_\mathrm{s} d)$, where $k_\mathrm{B}$ is the Boltzmann constant, $T$ sets the temperature, $\eta_\mathrm{s}$ is the dynamic viscosity of the solvent medium, and $d$ denotes the diameter of our suspended object.
Assuming normal conditions in water ($T = \SI{293.15}{\kelvin}$, $\eta_\mathrm{s} \approx \SI{0.001}{\newton\second\per\meter\squared}$), an object of diameter $d \approx \SI{10}{\micro\meter}$ is subject to a bare diffusivity of $D_0 \approx \SI{0.04}{\micro\meter\square\per\second}$.
Further, taking the example of bacterial turbulence observed in suspensions of $\textit{Bacillus Subtilis}$~\cite{dunkel2013fluid}, the average velocity is of the order of $v_\mathrm{a} \approx \SI{10}{\micro\meter\per\second}$ and the advective length scale is of the order of the mean vortex size $\Lambda_\mathrm{a} \approx \SI{40}{\micro\meter}$~\cite{dunkel2013fluid}.
As a result, the Peclet number is determined as $\mathrm{Pe}\approx 10^4$, which means that diffusion is several orders of magnitude weaker than advective transport and thus can be ignored.

The motion of small finite-size particles in nonhomogenous flow is described by the Maxey--Riley equation~\cite{maxey1983equation,brandt2022particle}.
Assuming that the diameters $d$ of our objects are notably smaller than the length scale of the flow patterns and thus neglecting Fax\'en corrections~\cite{haller2008inertial}, the equations of motion can be significantly simplified. %
Under these assumptions, the evolution of position $\mathbf{X}_i$ and velocity $\mathbf{U}_i$ of object $i$ reads~\cite{haller2008inertial}
\begin{equation}
\label{eq: equation of motion passive particle}
\frac{\partial \mathbf{X}_i}{\partial t} = \mathbf{U}_i\, , \qquad \frac{\partial \mathbf{U}_i}{\partial t} = \frac{3R}{2} \bigg[ \frac{\partial \mathbf{v}}{\partial t}(\mathbf{X}_i) + \mathbf{v}(\mathbf{X}_i) \cdot \nabla \mathbf{v}(\mathbf{X}_i) \bigg]- \frac{R}{\mathrm{Stk}} \big[\mathbf{U}_i - \mathbf{v}(\mathbf{\mathbf{X}_i})\big]  + \bigg(1 - \frac{3R}{2}\bigg) \mathbf{g}\, ,
\end{equation}
where $\mathbf{g}$ denotes gravitational acceleration.
The factor $R$ depends on the densities of the carrier fluid, $\rho_\mathrm{s}$, and the objects, $\rho_\mathrm{p}$, and is given as $R = 2\rho_\mathrm{s}/(\rho_\mathrm{s} + 2\rho_\mathrm{p})$.
Equation~(\ref{eq: equation of motion passive particle}) further includes the Stokes number $\mathrm{Stk}$, which measures the characteristic time of the motion of our objects with respect to the characteristic time of the flow.
Employing the same scaling as in the main text, that is, using the critical mode $k_\mathrm{c}$ as an inverse length scale and $(k_\mathrm{c}^2\nu_0)^{-1}$ as the time scale, the Stokes number in the present system is calculated as
\begin{equation}
\mathrm{Stk} =  \frac{8\pi^2}{9}\frac{d^2}{\Lambda_\mathrm{c}^2}\, ,
\end{equation}
where we have used the definition of the characteristic length scale $\Lambda_\mathrm{c} = 2\pi/k_\mathrm{c}$.
In our numerical calculations we set $\mathrm{Stk} = 0.033$.
This corresponds to a diameter of $d \approx 0.06 \Lambda_\mathrm{c}$, implying that our objects are about an order of magnitude smaller than the characteristic scale of the vortex patterns.

From numerous experimental, numerical, and theoretical studies of particle-laden flow~\cite{balachandar2010turbulent,brandt2022particle} we know that small objects tend to accumulate in certain areas of the flow field when their density is different from that of the carrier fluid.
In particular, a lower density of the objects leads to aggregation within vortices, while denser objects are ejected from vortices~\cite{eaton1994preferential}.
We transfer this effect to our present system to spatially organize passive objects by self-supported regular vortex patterns of the active shear-thickening fluid.
To this end, we vary the ratio of densities as the key parameter, changing the value of $R$ in Eq.~(\ref{eq: equation of motion passive particle}).
As a result, for lighter objects, we find a flow-induced clustering at vortex centers, as demonstrated in Fig.~5(a) in the main text for $\rho_\mathrm{p} = 0.5\rho_\mathrm{s}$.
Conversely, heavier objects accumulate between vortices, which leads to regular grids of elevated density of the objects, as shown in Fig.~5(b) in the main text for $\rho_\mathrm{p} = 2\rho_\mathrm{s}$ at an intermediate time scale.
In the latter case, after longer times have passed, most objects are found in the centers between any four neighboring vortices forming a rectangular structure.
Our motion is restricted to planar motion due to the geometrical confinement to thin films, sheets, or layers, which limits the influence of gravity in Eq.~(\ref{eq: equation of motion passive particle}).


%